\documentclass{article}

\usepackage[english]{babel}

\usepackage[dvips]{graphicx}

\usepackage[a4paper,top=2cm,bottom=2cm,left=3cm,right=3cm,marginparwidth=1.75cm]{geometry}

\usepackage{amsmath}
\usepackage{mathtools}
\usepackage{graphicx}
\usepackage[colorlinks=true, allcolors=blue]{hyperref}

\usepackage[braket, qm]{qcircuit}

\title{Entanglement Distance of Two- and Multi-Qubit Variational States and Its Quantification with Quantum Computing}

\begin{document}
\maketitle

\centerline {Kh. P. Gnatenko $^{1,2}$ \footnote{E-Mail address: khrystyna.gnatenko@gmail.com}, R. O. Hredil $^{1}$ , V. Y. Pinchuk $^1$ , M. Z. Seniak $^1$ , Y. T. Shevchuk$^1$}
\medskip

\centerline {\small  $^1$ \it Ivan Franko National University of Lviv,}
\centerline {\small \it Professor Ivan Vakarchuk Department for Theoretical Physics,}
\centerline {\small \it 12 Drahomanov St., Lviv, 79005, Ukraine}

\centerline{\small \it $^2$ SoftServe Inc.}

\begin{abstract}
We study the entanglement distance of variational quantum states for two-qubit and multi-qubit systems. These states are constructed using variational quantum circuits with $R_Y$ rotations and entangling $CZ$ gates. For the two-qubit case, we analytically derive recurrence relations for expectation values of Pauli observables. This approach allows us to  calculate quantum correlators and evaluate the entanglement distance as a function of the circuit parameters and depth. The analysis was extended to a closed one-dimensional chain of $N$ qubits. An explicit analytical expression for the entanglement is derived for the case of two layers. We conclude that the entanglement of a qubit with the rest of the system depends on the parameters of the gates acting on first- and second-nearest neighbors in the chain topology of the entangling layers. We also quantify the entanglement of the variational quantum states using quantum computing on the AerSimulator. The corresponding quantum protocols are constructed, and the dependence of the entanglement on the parameters of the variational quantum states is studied. The results of the quantum programming are in good agreement with the theoretical predictions.
\end{abstract}
\centerline{Key words: variational quantum states, entanglement distance, quantum computing}

\section{Introduction}

Entanglement is one of the fundamental features of quantum mechanics and serves as a key resource in quantum information processing, quantum computing, and quantum communication \cite{horodecki2009quantum, amico2008entanglement,Lloyd1996}. It plays a central role in protocols such as quantum teleportation, quantum cryptography, and in achieving quantum advantage in computational tasks. Therefore, the characterization and quantification of entanglement in multi-qubit quantum systems remain a pivotal challenge in modern quantum physics \cite{horodecki2009quantum,shimony1995degree, Horodecki2002Direct,Llewellyn2020Chip}.

Recently, the properties of multi-qubit quantum systems, including variational quantum states and spin models  have attracted considerable attention (see, for instance, \cite{dbnd-sl4j, PhysRevLett.134.050201,PRXQuantum.5.020355, Kuzmak2023EPL,gnatenko2025relation} and references therein) using specialized quantum programming frameworks \cite{Qiskit}. Specifically, the geometric properties of quantum graph states and their underlying topological structures have been investigated to reveal how particle correlations relate to network parameters (see, for instance, \cite{gnatenko2025relation,gnatenko2024entanglement} and references therein). This aligns with a broad international effort to investigate the expressivity and entangling capability of parameterized quantum circuits (PQCs) \cite{sim2019expressibility}. In particular, the hardware-efficient ansatz (HEA) has emerged as a crucial tool for near-term noisy intermediate-scale quantum (NISQ) devices, where structure optimization is vital for performance \cite{ostaszewski2021structure, joch2025entanglement, Cerezo2021, Rudolph2023}.

Among various approaches to quantifying entanglement, the entanglement distance (ED) provides a convenient  way to quantify how strongly a given qubit is correlated with the rest of the system \cite{Cocchiarella2020Entanglement}. This quantity can be expressed in terms of expectation values of Pauli operators. Consequently, it can be directly evaluated with quantum computing. Recent research has focused on both quantum and classical calculations of the entanglement distance and related geometric properties \cite{DeSimone2026}.

In this work, we study the entanglement distance of quantum states prepared using $R_Y$ rotations and $CZ$ gates. We begin with the two-qubit case, where we derive exact analytical expressions for expectation values of Pauli observables. We then extend the analysis to a multi-qubit system arranged in a closed one-dimensional chain. For this case, we derive analytical expressions for expectation values and entanglement distance for two layers. It is shown that with increasing circuit depth, more qubits influence a given qubit, reflecting the phenomenon of correlation spreading and entanglement growth in structured circuits \cite{nahum2017quantum, lieb1972finite}. Finally, we verify our analytical results using quantum computing. 

The paper is organized as follows. In Section \ref{sec:two-qubit} we derive analytical expressions for the entanglement of variational quantum states  in the two-qubit case. In Section \ref{sec:multi-qubit 1D chain} we consider multi-qubit variational quantum state  with entangled blocks of chain topology and examine dependence of the entanglement on the parameters of the variational quantum protocol. In Section \ref{sec: Simulation} we present results of quantum computing of the entanglement distance, and conclusions are given in Section \ref{sec: Conclusions}.

\section{Entanglement  of two-qubit variational quantum states and quantum correlators}
\label{sec:two-qubit}

To explore the properties of two-qubit variational quantum states, we consider a hardware-efficient parameterized quantum circuit (PQC) ansatz acting on the initial state $\ket{\psi_0} = \ket{00}$.

\begin{figure}[htpb]
    \centering
    \scalebox{1.0}{
    \Qcircuit @C=1.0em @R=0.2em @!R { \\
        \nghost{{q}_{0} :  } & \lstick{{q}_{0} :  } & \gate{\mathrm{R_Y}(\theta_0^{(1)})} & \ctrl{1} \barrier[0em]{1} & \cds{0}{\raisebox{3.5em}{$\ket{\psi_1}$}} & \gate{\mathrm{R_Y}(\theta_0^{(2)})} & \ctrl{1} \barrier[0em]{1} & \cds{0}{\raisebox{3.5em}{$\ket{\psi_2}$}} & \gate{\mathrm{R_Y}(\theta_0^{(3)})} & \ctrl{1} \barrier[0em]{1} & \cds{0}{\raisebox{3.5em}{$\ket{\psi_3}$}} & \qw & \cdots \\
        \nghost{{q}_{1} :  } & \lstick{{q}_{1} :  } & \gate{\mathrm{R_Y}(\theta_1^{(1)})} & \control\qw & \qw & \gate{\mathrm{R_Y}(\theta_1^{(2)})} & \control\qw & \qw & \gate{\mathrm{R_Y}(\theta_1^{(3)})} & \control\qw & \qw & \qw & \cdots \\
    \\ }}
    \vspace{0cm} 
    \caption{Periodic layered architecture of the two-qubit parameterized quantum circuit ansatz. Each single layer consists of independent $R_Y$ rotations followed by a $CZ$ entangling gate, defining the state $\ket{\psi_n}$ at depth $n$.}
    \label{fig:pqc_ansatz}
\end{figure}
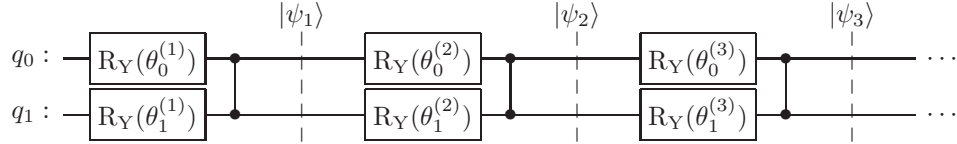
Denoting by $U_n$ the unitary corresponding to the $n$-th layer,
\begin{equation} \label{eq:u_layer_operator}
    U_n
    =
    CZ_{01} R_Y(\theta_0^{(n)}) R_Y(\theta_1^{(n)}),
\end{equation}
the state evolves according to
\begin{equation} \label{eq:state_evol_U}
    \ket{\psi_n} = U_n \ket{\psi_{n-1}}.
\end{equation}
This layered structure allows systematic scaling of circuit depth and expressivity.


To quantify bipartite entanglement in the considered two-qubit variational states, we use the entanglement distance introduced in \cite{Cocchiarella2020Entanglement}. For a pure two-qubit state, this quantity is determined from the reduced single-qubit state and can be expressed through the local Bloch vector components as

\begin{equation}
\label{ent dist}
E^{ED} = 1 - \langle X \rangle^2 -\langle Y \rangle^2 - \langle Z \rangle^2.
\end{equation}

Here, $\langle X \rangle$, $\langle Y \rangle$, and $\langle Z \rangle$ are expectation values of the Pauli operators for either qubit. For separable product states, the reduced state remains pure and $E^{ED}=0$, while nonzero values indicate entanglement between the qubits. Larger values correspond to stronger bipartite quantum correlations.

For a general observable $A$, we have
   $ \langle A \rangle_n
    =
    \bra{\psi_n} A \ket{\psi_n}.$
Using the recursive definition of $\ket{\psi_n}$ from \eqref{eq:u_layer_operator}-\eqref{eq:state_evol_U}, we rewrite this as:

\begin{equation}
    \langle A \rangle_n
    =
    \bra{\psi_{n-1}}
    \left(
    R_Y^{\dagger}(\theta_0^{(n)})
    R_Y^{\dagger}(\theta_1^{(n)})
    CZ_{01}^\dagger
    \; A \;
    CZ_{01}
    R_Y(\theta_0^{(n)})
    R_Y(\theta_1^{(n)})
    \right)
    \ket{\psi_{n-1}}.
\end{equation}
We compute this in two steps.


The $CZ$ gate is a Clifford operator, therefore it maps Pauli strings onto Pauli strings under conjugation. Its action on single-qubit Paulis is:

\begin{equation}\label{eq:cz_single_qubit}
\begin{aligned}
    CZ^\dagger X_0 CZ &= X_0 Z_1, &\qquad CZ^\dagger X_1 CZ &= Z_0 X_1, \\
    CZ^\dagger Y_0 CZ &= Y_0 Z_1, &\qquad CZ^\dagger Y_1 CZ &= Z_0 Y_1, \\
    CZ^\dagger Z_0 CZ &= Z_0,     &\qquad CZ^\dagger Z_1 CZ &= Z_1.
\end{aligned}
\end{equation}
From this we obtain all two-qubit transformations:

\begin{equation}\label{eq:cz_two_qubit}
\begin{aligned}
    CZ^\dagger (X_0 X_1) CZ &= Y_0 Y_1,     &\qquad CZ^\dagger (Y_0 X_1) CZ &= - X_0 Y_1, &\qquad CZ^\dagger (Z_0 X_1) CZ &= X_1, \\
    CZ^\dagger (X_0 Y_1) CZ &= - Y_0 X_1,  &\qquad CZ^\dagger (Y_0 Y_1) CZ &= X_0 X_1,  &\qquad CZ^\dagger (Z_0 Y_1) CZ &= Y_1, \\
    CZ^\dagger (X_0 Z_1) CZ &= X_0,        &\qquad CZ^\dagger (Y_0 Z_1) CZ &= Y_0,        &\qquad CZ^\dagger (Z_0 Z_1) CZ &= Z_0 Z_1.
\end{aligned}
\end{equation}
Thus any Pauli string after conjugation remains a tensor product of Pauli matrices.


The parameterized rotations in our ansatz, $R_Y(\theta_0^{(n)}) \otimes R_Y(\theta_1^{(n)})$, act independently on each qubit. Because these rotations are local, we only need to characterize their effect on a single-qubit Pauli basis. Under a rotation $R_Y(\theta) = \exp(-i \theta Y / 2)$, the Pauli matrices transform as:

\begin{equation}\label{eq:ry_transformation}
\begin{aligned}
    R_Y^\dagger(\theta) X R_Y(\theta)
    &=
    \cos\theta\, X + \sin\theta\, Z, \\
    R_Y^\dagger(\theta) Y R_Y(\theta)
    &=
    Y, \\
    R_Y^\dagger(\theta) Z R_Y(\theta)
    &=
    \cos\theta\, Z - \sin\theta\, X.
\end{aligned}
\end{equation}


By combining the conjugation steps described in Step 1 and Step 2, we can systematically derive recurrence relations for any observable. The general procedure involves evolving the operator $A$ backward through one layer of the circuit.

For instance, consider the expectation value $\langle X_0 \rangle_n$. Following the transformation of the operator $X_0$ using the $CZ$ rules from \eqref{eq:cz_single_qubit} and the rotation rules from \eqref{eq:ry_transformation}:
\begin{equation}
    X_0 \xrightarrow{CZ} X_0 Z_1 \xrightarrow{R_{Y_0} \otimes R_{Y_1}} (\cos\theta_0^{(n)} X_0 + \sin\theta_0^{(n)} Z_0) \otimes (\cos\theta_1^{(n)} Z_1 - \sin\theta_1^{(n)} X_1).
\end{equation}
Taking the expectation value with respect to $\ket{\psi_{n-1}}$ yields the first relation in \eqref{eq:single_expectations}. Analogously, by applying this logic to the remaining Pauli strings and utilizing the two-qubit transformations in \eqref{eq:cz_two_qubit}, we obtain the following complete set of recurrence relations.


The single-qubit expectation values evolve as:
\begin{equation} \label{eq:single_expectations}
\begin{aligned} 
\langle X_0 \rangle_n
    &= \cos\theta_0^{(n)}\cos\theta_1^{(n)} \langle X_0 Z_1 \rangle_{n-1} - \cos\theta_0^{(n)}\sin\theta_1^{(n)} \langle X_0 X_1 \rangle_{n-1} \\
    &+ \sin\theta_0^{(n)}\cos\theta_1^{(n)} \langle Z_0 Z_1 \rangle_{n-1} - \sin\theta_0^{(n)}\sin\theta_1^{(n)} \langle Z_0 X_1 \rangle_{n-1}, \\[10pt]
\langle Y_0 \rangle_n
    &= \cos \theta_1^{(n)} \langle Y_0 Z_1 \rangle_{n-1} - \sin \theta_1^{(n)} \langle Y_0 X_1 \rangle_{n-1}, \\[10pt]
\langle Z_0 \rangle_n
    &= \cos \theta_0^{(n)} \langle Z_0 \rangle_{n-1} - \sin \theta_0^{(n)} \langle X_0 \rangle_{n-1}, \\[20pt]
\langle X_1 \rangle_n
    &= \cos\theta_0^{(n)}\cos\theta_1^{(n)} \langle Z_0 X_1 \rangle_{n-1} + \cos\theta_0^{(n)}\sin\theta_1^{(n)} \langle Z_0 Z_1 \rangle_{n-1} \\
    &- \sin\theta_0^{(n)}\cos\theta_1^{(n)} \langle X_0 X_1 \rangle_{n-1} - \sin\theta_0^{(n)}\sin\theta_1^{(n)} \langle X_0 Z_1 \rangle_{n-1}, \\[10pt]
\langle Y_1 \rangle_n
    &= \cos \theta_0^{(n)} \langle X_0 Y_1 \rangle_{n-1} + \sin \theta_0^{(n)} \langle Z_0 Y_1 \rangle_{n-1}, \\[10pt]
\langle Z_1 \rangle_n
    &= \cos \theta_1^{(n)} \langle Z_1 \rangle_{n-1} - \sin \theta_1^{(n)} \langle X_1 \rangle_{n-1}, \\[10pt]
\end{aligned}
\end{equation}


The correlations between qubits follow more complex paths due to the entangling $CZ$ gate are as follows

\begin{equation} \label{eq:2qubit_expectations}
\begin{aligned} 
    \langle X_0 X_1 \rangle_n
    &= \langle Y_0 Y_1 \rangle_{n-1}, \\[10pt]
    \langle X_0 Y_1 \rangle_n
    &= -\cos \theta_1^{(n)} \langle Y_0 X_1 \rangle_{n-1} - \sin \theta_1^{(n)} \langle Y_0 Z_1 \rangle_{n-1}, \\[10pt]
\langle X_0 Z_1 \rangle_n
    &= \cos \theta_0^{(n)} \langle X_0 \rangle_{n-1} + \sin \theta_0^{(n)} \langle Z_0 \rangle_{n-1}, \\[20pt]
    \langle Y_0 X_1 \rangle_n
    &= -\cos \theta_0^{(n)} \langle X_0 Y_1 \rangle_{n-1} - \sin \theta_0^{(n)} \langle Z_0 Y_1 \rangle_{n-1}, \\[10pt]
    \langle Y_0 Y_1 \rangle_n
    &= \cos\theta_0^{(n)}\cos\theta_1^{(n)} \langle X_0 X_1 \rangle_{n-1} + \cos\theta_0^{(n)}\sin\theta_1^{(n)} \langle X_0 Z_1 \rangle_{n-1} \\
    &+ \sin\theta_0^{(n)}\cos\theta_1^{(n)} \langle Z_0 X_1 \rangle_{n-1} + \sin\theta_0^{(n)}\sin\theta_1^{(n)} \langle Z_0 Z_1 \rangle_{n-1}, \\[10pt]
    \langle Y_0 Z_1 \rangle_n
    &= \langle Y_0 \rangle_{n-1}, \\[20pt]
    \langle Z_0 X_1 \rangle_n
    &= \cos \theta_1^{(n)} \langle X_1 \rangle_{n-1} + \sin \theta_1^{(n)} \langle Z_1 \rangle_{n-1}, \\[10pt]
    \langle Z_0 Y_1 \rangle_n
    &= \langle Y_1 \rangle_{n-1}, \\[10pt]
    \langle Z_0 Z_1 \rangle_n
    &= \cos\theta_0^{(n)}\cos\theta_1^{(n)} \langle Z_0 Z_1 \rangle_{n-1} - \cos\theta_0^{(n)}\sin\theta_1^{(n)} \langle Z_0 X_1 \rangle_{n-1} \\
    &- \sin\theta_0^{(n)}\cos\theta_1^{(n)} \langle X_0 Z_1 \rangle_{n-1} + \sin\theta_0^{(n)}\sin\theta_1^{(n)} \langle X_0 X_1 \rangle_{n-1}.
\end{aligned}
\end{equation}

Using the general representation of the resulting state vector $\ket{\psi_n} = a_n\ket{00} + b_n\ket{01} + c_n\ket{10} + d_n\ket{11}$, where all coefficients are real-valued, we obtain the simplified formula for the entanglement distance:

\begin{equation}\label{eq:E_layer_n}
\begin{aligned}
E_n^{ED} &= 1 - \left( \langle X_0 \rangle_n^2 + \langle Y_0 \rangle_n^2 + \langle Z_0 \rangle_n^2 \right) =
(a_n^2 + b_n^2 + c_n^2 + d_n^2)^2 - 4(a_n c_n + b_n d_n)^2\\
&\quad - (a_n^2 + b_n^2 - c_n^2 - d_n^2)^2 = 4(b_n c_n - a_n d_n)^2 = \langle Y_0 Y_1 \rangle_n^2
\end{aligned}
\end{equation}

Noticeably, the entanglement distance of a two-qubit variational quantum state perfectly matches the square of the respective two-qubit correlator $\langle Y_0 Y_1 \rangle$.

\subsection{Entanglement of one- and two-layer variational quantum states of two qubits}

Before we can use the recurrence relations derived above, we define the base case by directly evaluating the expectation values for the initial state $\ket{\psi_0} = \ket{00}$. The single-qubit expectation values for the initial state are:

\begin{equation}\label{eq:base_case_single_qubit}
\begin{aligned}
    \langle X_0 \rangle_0 &= 0, &\qquad \langle X_1 \rangle_0 &= 0, \\
    \langle Y_0 \rangle_0 &= 0, &\qquad \langle Y_1 \rangle_0 &= 0, \\
    \langle Z_0 \rangle_0 &= 1, &\qquad \langle Z_1 \rangle_0 &= 1.
\end{aligned}
\end{equation}

The two-qubit correlations for the initial state have the following expectation values:

\begin{equation}\label{eq:base_case_two_qubit}
\begin{aligned}
    \langle X_0 X_1 \rangle_0 &= 0, &\qquad \langle Y_0 X_1 \rangle_0 &= 0, &\qquad \langle Z_0 X_1 \rangle_0 &= 0, \\
    \langle X_0 Y_1 \rangle_0 &= 0, &\qquad \langle Y_0 Y_1 \rangle_0 &= 0, &\qquad \langle Z_0 Y_1 \rangle_0 &= 0, \\
    \langle X_0 Z_1 \rangle_0 &= 0, &\qquad \langle Y_0 Z_1 \rangle_0 &= 0, &\qquad \langle Z_0 Z_1 \rangle_0 &= 1.
\end{aligned}
\end{equation}

Now we can apply the recurrence relations to the base case:

\begin{equation}\label{eq:layer1_single_qubit}
\begin{aligned}
    \langle X_0 \rangle_1 &= \sin\theta_0^{(1)}\cos\theta_1^{(1)}, &\qquad \langle X_1 \rangle_1 &= \cos\theta_0^{(1)}\sin\theta_1^{(1)}, \\
    \langle Y_0 \rangle_1 &= 0, &\qquad \langle Y_1 \rangle_1 &= 0, \\
    \langle Z_0 \rangle_1 &= \cos\theta_0^{(1)}, &\qquad \langle Z_1 \rangle_1 &= \cos\theta_1^{(1)};
\end{aligned}
\end{equation}

\begin{equation}\label{eq:layer1_two_qubit}
\begin{aligned}
    \langle X_0 X_1 \rangle_1 &= 0, &\qquad \langle Y_0 X_1 \rangle_1 &= 0, &\qquad \langle Z_0 X_1 \rangle_1 &= \sin\theta_1^{(1)}, \\
    \langle X_0 Y_1 \rangle_1 &= 0, &\qquad \langle Y_0 Y_1 \rangle_1 &= \sin\theta_0^{(1)}\sin\theta_1^{(1)}, &\qquad \langle Z_0 Y_1 \rangle_1 &= 0, \\
    \langle X_0 Z_1 \rangle_1 &= \sin\theta_0^{(1)}, &\qquad \langle Y_0 Z_1 \rangle_1 &= 0, &\qquad \langle Z_0 Z_1 \rangle_1 &= \cos\theta_0^{(1)}\cos\theta_1^{(1)}.
\end{aligned}
\end{equation}

By applying the recurrence relations one more time we get the expectation values for a two-layer circuit:

\begin{equation} \label{eq:layer2_single_qubit}
\begin{aligned} 
\langle X_0 \rangle_2
    &= \sin\theta_0^{(1)}\cos\theta_0^{(2)}\cos\theta_1^{(2)} + \cos\theta_0^{(1)}\cos\theta_1^{(1)}\sin\theta_0^{(2)}\cos\theta_1^{(2)} \\
    &- \sin\theta_1^{(1)}\sin\theta_0^{(2)}\sin\theta_1^{(2)}, \\[10pt]
\langle Y_0 \rangle_2
    &= 0, \\[10pt]
\langle Z_0 \rangle_2
    &= \cos\theta_0^{(1)}\cos\theta_0^{(2)} - \sin\theta_0^{(1)}\cos\theta_1^{(1)}\sin\theta_0^{(2)}, \\[20pt]
\langle X_1 \rangle_2
    &= \sin\theta_1^{(1)}\cos\theta_0^{(2)}\cos\theta_1^{(2)} + \cos\theta_0^{(1)}\cos\theta_1^{(1)}\cos\theta_0^{(2)}\sin\theta_1^{(2)} \\
    &- \sin\theta_0^{(1)}\sin\theta_0^{(2)}\sin\theta_1^{(2)}, \\[10pt]
\langle Y_1 \rangle_2
    &= 0, \\[10pt]
\langle Z_1 \rangle_2
    &= \cos\theta_1^{(1)}\cos\theta_1^{(2)} - \cos\theta_0^{(1)}\sin\theta_1^{(1)}\sin\theta_1^{(2)}, \\[20pt]
\end{aligned}
\end{equation}

\begin{equation} \label{eq:layer2_two_qubit}
\begin{aligned} 
    \langle X_0 X_1 \rangle_2
    &= \sin\theta_0^{(1)}\sin\theta_1^{(1)}, \\[10pt]
    \langle X_0 Y_1 \rangle_2
    &= 0, \\[10pt]
    \langle X_0 Z_1 \rangle_2
    &= \sin\theta_0^{(1)}\cos\theta_1^{(1)}\cos\theta_0^{(2)} + \cos\theta_0^{(1)}\sin\theta_0^{(2)}, \\[20pt]
    \langle Y_0 X_1 \rangle_2
    &= 0, \\[10pt]
    \langle Y_0 Y_1 \rangle_2
    &= \sin\theta_0^{(1)}\cos\theta_0^{(2)}\sin\theta_1^{(2)} + \sin\theta_1^{(1)}\sin\theta_0^{(2)}\cos\theta_1^{(2)} \\
    &+ \cos\theta_0^{(1)}\cos\theta_1^{(1)}\sin\theta_0^{(2)}\sin\theta_1^{(2)}, \\[10pt]
    \langle Y_0 Z_1 \rangle_2
    &= 0, \\[20pt]
    \langle Z_0 X_1 \rangle_2
    &= \cos\theta_0^{(1)}\sin\theta_1^{(1)}\cos\theta_1^{(2)} + \cos\theta_1^{(1)}\sin\theta_1^{(2)}, \\[10pt]
    \langle Z_0 Y_1 \rangle_2
    &= 0, \\[10pt]
    \langle Z_0 Z_1 \rangle_2
    &= \cos\theta_0^{(1)}\cos\theta_1^{(1)}\cos\theta_0^{(2)}\cos\theta_1^{(2)} - \sin\theta_1^{(1)}\cos\theta_0^{(2)}\sin\theta_1^{(2)} \\
    &- \sin\theta_0^{(1)}\sin\theta_0^{(2)}\cos\theta_1^{(2)}.
\end{aligned}
\end{equation}

We can apply this procedure as many times as necessary. The single-qubit expectation values for a three-layer circuit are as follows:

\begin{equation} \label{eq:layer3_single_qubit}
\begin{aligned} 
\langle X_0 \rangle_3
    &= (\sin\theta_0^{(1)}\cos\theta_1^{(1)}\cos\theta_0^{(2)} + \cos\theta_0^{(1)}\sin\theta_0^{(2)})\cos\theta_0^{(3)}\cos\theta_1^{(3)} - \sin\theta_0^{(1)}\sin\theta_1^{(1)}\cos\theta_0^{(3)}\sin\theta_1^{(3)} \\
    &+ (\cos\theta_0^{(1)}\cos\theta_1^{(1)}\cos\theta_0^{(2)}\cos\theta_1^{(2)} - \sin\theta_1^{(1)}\cos\theta_0^{(2)}\sin\theta_1^{(2)} - \sin\theta_0^{(1)}\sin\theta_0^{(2)}\cos\theta_1^{(2)})\sin\theta_0^{(3)}\cos\theta_1^{(3)} \\
    &- (\cos\theta_0^{(1)}\sin\theta_1^{(1)}\cos\theta_1^{(2)} + \cos\theta_1^{(1)}\sin\theta_1^{(2)})\sin\theta_0^{(3)}\sin\theta_1^{(3)}, \\[10pt]
\langle Y_0 \rangle_3
    &= 0, \\[10pt]
\langle Z_0 \rangle_3
    &= (\cos\theta_0^{(1)}\cos\theta_0^{(2)} - \sin\theta_0^{(1)}\cos\theta_1^{(1)}\sin\theta_0^{(2)})\cos\theta_0^{(3)} \\
    &- (\sin\theta_0^{(1)}\cos\theta_0^{(2)}\cos\theta_1^{(2)} + \cos\theta_0^{(1)}\cos\theta_1^{(1)}\sin\theta_0^{(2)}\cos\theta_1^{(2)} - \sin\theta_1^{(1)}\sin\theta_0^{(2)}\sin\theta_1^{(2)})\sin\theta_0^{(3)}, \\[20pt]
\langle X_1 \rangle_3
    &= (\cos\theta_0^{(1)}\sin\theta_1^{(1)}\cos\theta_1^{(2)} + \cos\theta_1^{(1)}\sin\theta_1^{(2)})\cos\theta_0^{(3)}\cos\theta_1^{(3)} \\
    &+ (\cos\theta_0^{(1)}\cos\theta_1^{(1)}\cos\theta_0^{(2)}\cos\theta_1^{(2)} - \sin\theta_1^{(1)}\cos\theta_0^{(2)}\sin\theta_1^{(2)} - \sin\theta_0^{(1)}\sin\theta_0^{(2)}\cos\theta_1^{(2)})\cos\theta_0^{(3)}\sin\theta_1^{(3)} \\
    &- \sin\theta_0^{(1)}\sin\theta_1^{(1)}\sin\theta_0^{(3)}\cos\theta_1^{(3)} - (\sin\theta_0^{(1)}\cos\theta_1^{(1)}\cos\theta_0^{(2)} + \cos\theta_0^{(1)}\sin\theta_0^{(2)})\sin\theta_0^{(3)}\sin\theta_1^{(3)}, \\[10pt]
\langle Y_1 \rangle_3
    &= 0, \\[10pt]
\langle Z_1 \rangle_3
    &= (\cos\theta_1^{(1)}\cos\theta_1^{(2)} - \cos\theta_0^{(1)}\sin\theta_1^{(1)}\sin\theta_1^{(2)})\cos\theta_1^{(3)} \\
    &- (\sin\theta_1^{(1)}\cos\theta_0^{(2)}\cos\theta_1^{(2)} + \cos\theta_0^{(1)}\cos\theta_1^{(1)}\cos\theta_0^{(2)}\sin\theta_1^{(2)})\sin\theta_1^{(3)}. \\
\end{aligned}
\end{equation}


Then, using the definition and the obtained expectation values, the entanglement distance—which is identical whether calculated for the first or the second qubit due to the system's symmetric structure—is given by:

\begin{equation}\label{eq:E_layer1}
E^{ED}_1 = \langle Y_0 Y_1 \rangle_1^2 = \sin^2 \theta_0^{(1)}\sin^2 \theta_1^{(1)},
\end{equation}

\begin{equation}\label{eq:E_layer2}
\begin{aligned}
    E_2^{ED} =  \langle Y_0 Y_1 \rangle_2^2 &= \Biggl( \sin\theta_0^{(1)}\cos\theta_0^{(2)}\sin\theta_1^{(2)} + \cos\theta_0^{(1)}\cos\theta_1^{(1)}\sin\theta_0^{(2)}\sin\theta_1^{(2)} + \\
    &\quad + \sin\theta_1^{(1)}\sin\theta_0^{(2)}\cos\theta_1^{(2)} \Biggr)^2.
\end{aligned}
\end{equation}

Using the analytical expressions for expectation values derived in the previous subsections, we evaluate $E^{ED}$ for different circuit parameters and compare the results with numerical simulations in the following section.

\subsection{Quantum computing of the entanglement and comparison with analytical results of the two-qubit case}

To verify the analytical expressions obtained for the two-qubit case, we performed numerical simulations using the \texttt{qiskit\_aer} simulator. We considered a two-qubit circuit with three layers of $R_Y$ rotations and $CZ$ gates.

The entanglement distance was evaluated for different choices of circuit parameters. In particular, we considered three cases:
\begin{itemize}
    \item[(a)] $\theta^{(1)} = \pi/2$, varying $\theta^{(2)}$ and $\theta^{(3)}$,
    \item[(b)] $\theta^{(2)} = \pi/2$, varying $\theta^{(1)}$ and $\theta^{(3)}$,
    \item[(c)] $\theta^{(3)} = \pi/2$, varying $\theta^{(1)}$ and $\theta^{(2)}$.
\end{itemize}

\begin{figure}[htpb]
    \centering
    \includegraphics[width=0.6\textwidth]{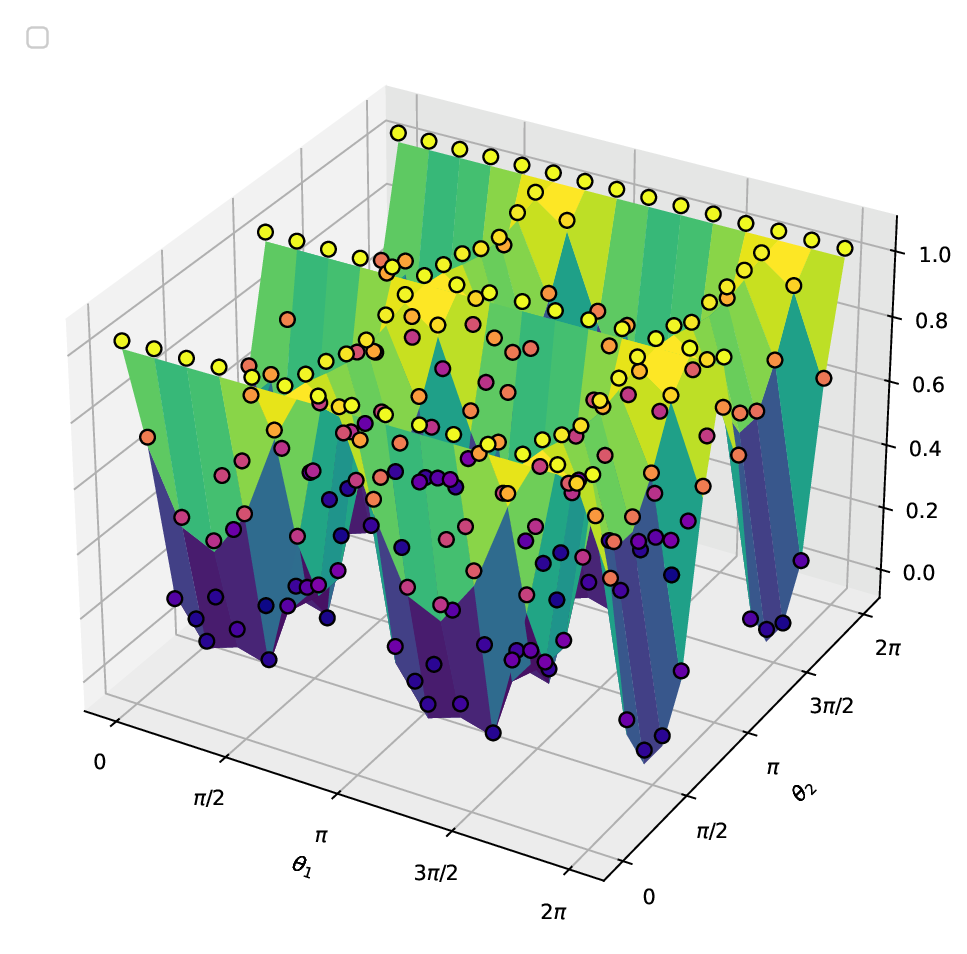}
    \caption{Surface plot of the entanglement distance \\$E^{ED} = \left(\cos\theta_0^{(3)}\cos\theta_1^{(3)} - \left(\cos\theta_0^{(2)}\sin\theta_1^{(2)} + \sin\theta_0^{(2)}\cos\theta_1^{(2)}\right) \sin\theta_0^{(3)}\sin\theta_1^{(3)}\right)^2$ for $\theta^{(1)} = \pi/2$ as a function of $\theta^{(2)}$ and $\theta^{(3)}$. The semi-transparent surface represents the analytical result, while the points correspond to numerical simulations.}
    \label{fig:theta1}
\end{figure}
\begin{figure}[htpb]
    \centering
    \includegraphics[width=0.6\textwidth]{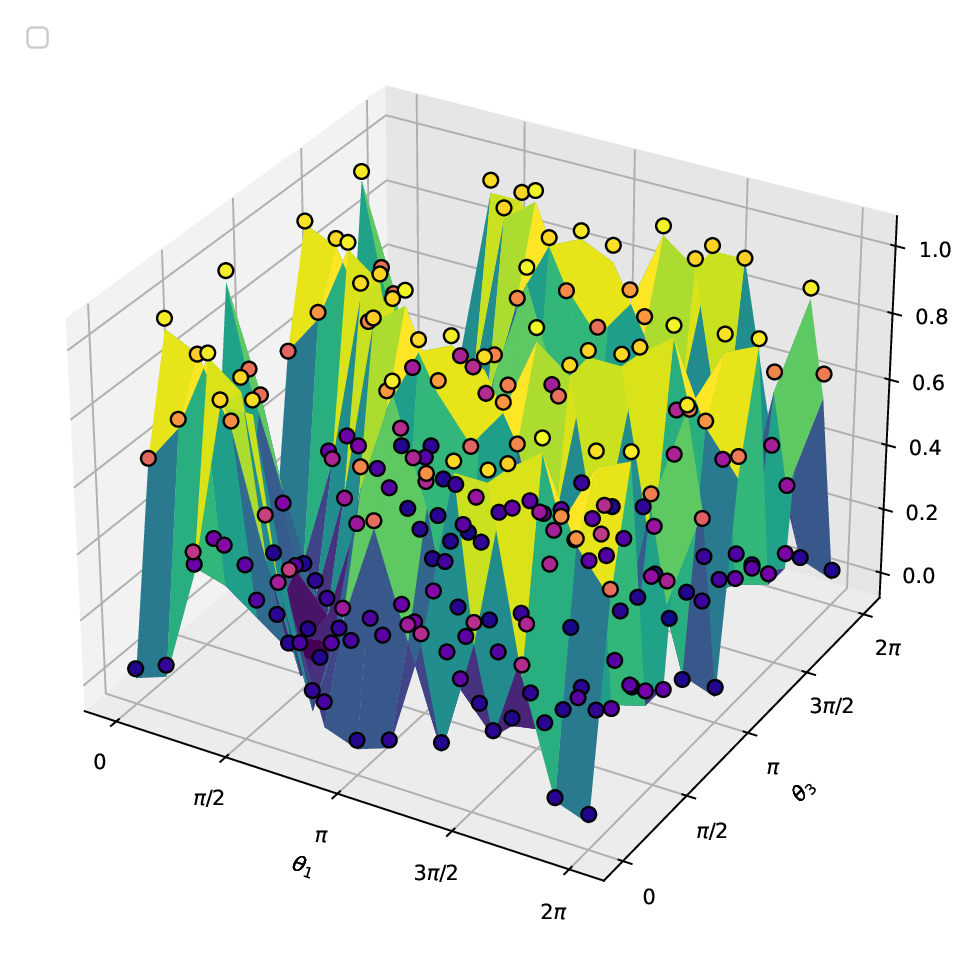}
    \caption{Surface plot of the entanglement distance \\$E^{ED} = \left(\sin\theta_0^{(1)}\sin\theta_1^{(1)}\cos\theta_0^{(3)}\cos\theta_1^{(3)} + \cos\theta_0^{(1)}\cos\theta_0^{(3)}\sin\theta_1^{(3)} + \cos\theta_1^{(1)}\sin\theta_0^{(3)}\cos\theta_1^{(3)}\right)^2$ for $\theta^{(2)} = \pi/2$ as a function of $\theta^{(1)}$ and $\theta^{(3)}$.}
    \label{fig:theta2}
\end{figure}
\begin{figure}[htpb]
    \centering
    \includegraphics[width=0.6\textwidth]{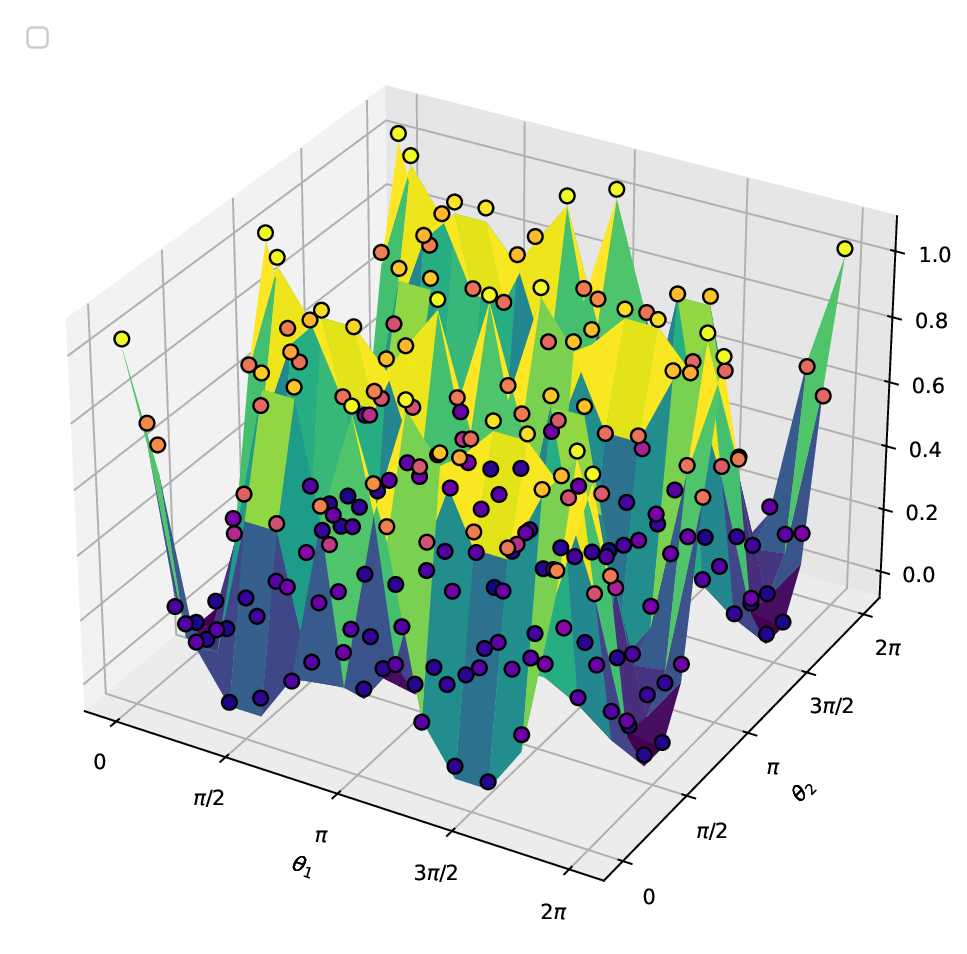}
    \caption{Surface plot of the entanglement distance \\$E^{ED} = \left(\cos\theta_0^{(1)}\cos\theta_1^{(1)}\cos\theta_0^{(2)}\cos\theta_1^{(2)} - \sin\theta_1^{(1)}\cos\theta_0^{(2)}\sin\theta_1^{(2)} - \sin\theta_0^{(1)}\sin\theta_0^{(2)}\cos\theta_1^{(2)}\right)^2$ for $\theta^{(3)} = \pi/2$ as a function of $\theta^{(1)}$ and $\theta^{(2)}$.}
    \label{fig:theta3}
\end{figure}

The numerical results were compared with the analytical predictions derived in the previous section. The comparison is presented in Figs.~\ref{fig:theta1}, \ref{fig:theta2}, and \ref{fig:theta3}. In these plots, the analytical results are shown as smooth surfaces, while the simulation results are represented by discrete points.

To quantify the agreement between analytical and numerical results, we evaluated the deviation between the two datasets. The obtained values for the three considered cases are:

\begin{itemize}
    \item[(a)] $\theta^{(1)}=\pi/2$: MAE $=0.019$, RMSE $=0.028$, correlation $=0.997$,
    \item[(b)] $\theta^{(2)}=\pi/2$: MAE $=0.022$, RMSE $=0.029$, correlation $=0.997$,
    \item[(c)] $\theta^{(3)}=\pi/2$: MAE $=0.024$, RMSE $=0.032$, correlation $=0.996$.
\end{itemize}

These results demonstrate an excellent agreement between analytical predictions and numerical simulations. The small discrepancies are due to statistical errors arising from the finite number of measurement shots.

\section{Entanglement properties of multi-qubit variational quantum states prepared with entangled blocks of chain topology}
\label{sec:multi-qubit 1D chain}

To generalize our analysis, we consider a closed one-dimensional chain of $N$ qubits subject to the same layered ansatz. We focus on an arbitrary target qubit $a$. For a closed chain topology, its nearest neighbors are $a-1$ and $a+1$ (with indices taken modulo $N$). We aim to compute the local expectation values $\langle X_a \rangle_n$, $\langle Y_a \rangle_n$, and $\langle Z_a \rangle_n$ in order to evaluate the entanglement distance of qubit $a$ with the rest of the system, defined in Eq.~(\ref{ent dist}).

 We denote the layer operation as $U_1 = C R_1$, where $C$ represents the layer of $CZ$ gates and $R_1$ the layer of $R_Y$ rotations.
\begin{equation}
\begin{aligned}
    \langle Z_a \rangle_1 &= \cos\theta_a^{(1)}, \\
    \langle Y_a \rangle_1 &= 0, \\
    \langle X_a \rangle_1 &= \sin\theta_a^{(1)} \cos\theta_{a-1}^{(1)} \cos\theta_{a+1}^{(1)}.
\end{aligned}
\end{equation}
The expectation value of $Y_a$ is strictly zero because the initial state and all transformations ($R_Y$ and $CZ$) are strictly real, forcing the expectation value of the purely imaginary Pauli $Y$ operator to vanish. This property holds for any depth $n$, meaning $\langle Y_a \rangle_n = 0 \ \forall n$.

With the single-qubit expectation values established, we can evaluate the entanglement distance for the target qubit $a$ after the first layer. Substituting the layer 1 values into Eq.~\eqref{ent dist}, we obtain:

\begin{equation} \label{eq:ED1_chain}
\begin{aligned}
    E^{ED}_1 &= 1 - \langle X_a \rangle_1^2 - \langle Y_a \rangle_1^2 - \langle Z_a \rangle_1^2 \\
    &= 1 - \left( \sin\theta_a^{(1)} \cos\theta_{a-1}^{(1)} \cos\theta_{a+1}^{(1)} \right)^2 - 0 - \left( \cos\theta_a^{(1)} \right)^2 \\
    &= \sin^2\theta_a^{(1)} - \sin^2\theta_a^{(1)} \cos^2\theta_{a-1}^{(1)} \cos^2\theta_{a+1}^{(1)} \\
    &= \sin^2\theta_a^{(1)} \left( 1 - \cos^2\theta_{a-1}^{(1)} \cos^2\theta_{a+1}^{(1)} \right).
\end{aligned}
\end{equation}

This simplified form elegantly illustrates the local generation of entanglement. The bipartite entanglement between qubit $a$ and the rest of the chain is gated entirely by its own local rotation $\sin^2\theta_a^{(1)}$ acting as a primary weight, while the term in the parentheses captures the essential loss of local purity due to the $CZ$ entangling operations with its immediate neighbors $a-1$ and $a+1$.

For the second layer ($n=2$), the state is given by $\ket{\psi_2} = U_2 \ket{\psi_1} = C R_2 \ket{\psi_1}$. We propagate the operators backward through $U_2$. For the $Z_a$ observable, since $CZ$ gates commute with $Z$, the operator only transforms under the local rotations:
\begin{equation}
    \langle Z_a \rangle_2 = \bra{\psi_1} R_2^\dagger C^\dagger Z_a C R_2 \ket{\psi_1} = \bra{\psi_1} R_2^\dagger Z_a R_2 \ket{\psi_1}.
\end{equation}
Substituting the first layer's expectation values, we get:
\begin{equation}
    \langle Z_a \rangle_2 = \cos\theta_a^{(2)} \langle Z_a \rangle_1 - \sin\theta_a^{(2)} \langle X_a \rangle_1 = \cos\theta_a^{(2)} \cos\theta_a^{(1)} - \sin\theta_a^{(2)} \sin\theta_a^{(1)} \cos\theta_{a-1}^{(1)} \cos\theta_{a+1}^{(1)}.
\end{equation}

The evolution of $X_a$ is more involved due to the entangling nature of the $CZ$ gates, which spreads the operator to the neighboring qubits. Passing $X_a$ through $U_2$ yields:
\begin{equation}
    U_2^\dagger X_a U_2 = R_2^\dagger C^\dagger X_a C R_2 = R_2^\dagger (X_a Z_{a-1} Z_{a+1}) R_2.
\end{equation}
Applying the local $R_Y$ rotations distributes the operators as follows:
\begin{equation}
    U_2^\dagger X_a U_2 = \left( X_a \cos\theta_a^{(2)} + Z_a \sin\theta_a^{(2)} \right) \left( Z_{a-1} \cos\theta_{a-1}^{(2)} - X_{a-1} \sin\theta_{a-1}^{(2)} \right) \left( Z_{a+1} \cos\theta_{a+1}^{(2)} - X_{a+1} \sin\theta_{a+1}^{(2)} \right).
\end{equation}

To find $\langle X_a \rangle_2$, this operator must be evaluated on the state $\ket{\psi_1} = C R_1 \ket{0}^{\otimes N}$. Therefore, we need to compute:
\begin{equation}
    \langle X_a \rangle_2 = \bra{0^{\otimes N}} R_1^\dagger C^\dagger \left[ U_2^\dagger X_a U_2 \right] C R_1 \ket{0^{\otimes N}}.
\end{equation}
We can evaluate this efficiently by passing the $C$ operation into the product of parentheses, using the distributive property $C^\dagger (A \cdot B \cdot D) C = (C^\dagger A C)(C^\dagger B C)(C^\dagger D C)$. Let us define the conjugated brackets as $B_1$, $B_2$, and $B_3$:
\begin{equation}
\begin{aligned}
    B_1 &= C^\dagger \left( X_a \cos\theta_a^{(2)} + Z_a \sin\theta_a^{(2)} \right) C = X_a Z_{a-1} Z_{a+1} \cos\theta_a^{(2)} + Z_a \sin\theta_a^{(2)} \equiv C_1 + S_1, \\
    B_2 &= C^\dagger \left( Z_{a-1} \cos\theta_{a-1}^{(2)} - X_{a-1} \sin\theta_{a-1}^{(2)} \right) C = Z_{a-1} \cos\theta_{a-1}^{(2)} - X_{a-1} Z_{a-2} Z_a \sin\theta_{a-1}^{(2)}, \\
    B_3 &= C^\dagger \left( Z_{a+1} \cos\theta_{a+1}^{(2)} - X_{a+1} \sin\theta_{a+1}^{(2)} \right) C = Z_{a+1} \cos\theta_{a+1}^{(2)} - X_{a+1} Z_a Z_{a+2} \sin\theta_{a+1}^{(2)}.
\end{aligned}
\end{equation}

The expectation value is then given by $\bra{0^{\otimes N}} R_1^\dagger V_x R_1 \ket{0^{\otimes N}}$, where $V_x = B_1 B_2 B_3$. Expanding the product of the neighboring terms $B_2$ and $B_3$, and utilizing the identity $Z_a^2 = I$, we obtain:
\begin{equation}
\begin{aligned}
    B_2 B_3 &= Z_{a-1} Z_{a+1} \cos\theta_{a-1}^{(2)} \cos\theta_{a+1}^{(2)} - Z_a Z_{a-1} X_{a+1} Z_{a+2} \cos\theta_{a-1}^{(2)} \sin\theta_{a+1}^{(2)} \\
    &\quad - Z_a X_{a-1} Z_{a-2} Z_{a+1} \sin\theta_{a-1}^{(2)} \cos\theta_{a+1}^{(2)} + X_{a-1} X_{a+1} Z_{a-2} Z_{a+2} \sin\theta_{a-1}^{(2)} \sin\theta_{a+1}^{(2)}.
\end{aligned}
\end{equation}

We evaluate the product $V_x = (C_1 + S_1) B_2 B_3$ by computing the expectation value of each resulting term over the initial state rotated by the first layer, i.e., $R_1 \ket{0}^{\otimes N}$. For such a state, the local single-qubit expectation values are $\langle Z_k \rangle_1 = \cos\theta_k^{(1)}$, $\langle X_k \rangle_1 = \sin\theta_k^{(1)}$, and $\langle Y_k \rangle_1 = 0$. Consequently, any product of operators that evaluates to a purely imaginary $Y$ Pauli matrix (e.g., $X \cdot Z = -iY$) on any qubit will vanish upon averaging.

When multiplying the term $C_1 = X_a Z_{a-1} Z_{a+1} \cos\theta_a^{(2)}$ with $B_2 B_3$, only the first term survives the averaging process, as the subsequent three generate $X_a Z_a \propto Y_a$ or $X_{a\pm1} Z_{a\pm1} \propto Y_{a\pm1}$, yielding zero. The single non-vanishing contribution is:
\begin{equation}
    \langle C_1 B_2 B_3 \rangle_1 = \langle X_a \rangle_1 \cos\theta_a^{(2)} \cos\theta_{a-1}^{(2)} \cos\theta_{a+1}^{(2)} = \sin\theta_a^{(1)} \cos\theta_a^{(2)} \cos\theta_{a-1}^{(2)} \cos\theta_{a+1}^{(2)}.
\end{equation}

Conversely, when multiplying $S_1 = Z_a \sin\theta_a^{(2)}$ with $B_2 B_3$, all four terms survive. The additional $Z_a$ operator squares to the identity when acting on the $Z_a$ components of $B_2 B_3$, and acts directly to yield $\langle Z_a \rangle_1$ elsewhere. Averaging these four components produces:
\begin{equation}
\begin{aligned}
    \langle S_1 B_2 B_3 \rangle_1 &= \langle Z_a Z_{a-1} Z_{a+1} \rangle_1 \sin\theta_a^{(2)} \cos\theta_{a-1}^{(2)} \cos\theta_{a+1}^{(2)} \\
    &\quad - \langle Z_{a-1} X_{a+1} Z_{a+2} \rangle_1 \sin\theta_a^{(2)} \cos\theta_{a-1}^{(2)} \sin\theta_{a+1}^{(2)} \\
    &\quad - \langle X_{a-1} Z_{a-2} Z_{a+1} \rangle_1 \sin\theta_a^{(2)} \sin\theta_{a-1}^{(2)} \cos\theta_{a+1}^{(2)} \\
    &\quad + \langle Z_a X_{a-1} X_{a+1} Z_{a-2} Z_{a+2} \rangle_1 \sin\theta_a^{(2)} \sin\theta_{a-1}^{(2)} \sin\theta_{a+1}^{(2)}.
\end{aligned}
\end{equation}

Substituting the independent single-qubit expectation values for layer 1, and summing the contributions from both $C_1$ and $S_1$, we finally arrive at the exact analytical expression for $\langle X_a \rangle_2$:
\begin{equation}
\begin{aligned}
    \langle X_a \rangle_2 &= \sin\theta_a^{(1)} \cos\theta_a^{(2)} \cos\theta_{a-1}^{(2)} \cos\theta_{a+1}^{(2)} \\
    &\quad + \sin\theta_a^{(2)} \Big[ \cos\theta_a^{(1)} \cos\theta_{a-1}^{(1)} \cos\theta_{a+1}^{(1)} \cos\theta_{a-1}^{(2)} \cos\theta_{a+1}^{(2)} - \cos\theta_{a-1}^{(1)} \sin\theta_{a+1}^{(1)} \cos\theta_{a+2}^{(1)} \cos\theta_{a-1}^{(2)} \sin\theta_{a+1}^{(2)} \\
    &\qquad\quad - \sin\theta_{a-1}^{(1)} \cos\theta_{a-2}^{(1)} \cos\theta_{a+1}^{(1)} \sin\theta_{a-1}^{(2)} \cos\theta_{a+1}^{(2)} \\
    &\qquad\quad + \cos\theta_a^{(1)} \sin\theta_{a-1}^{(1)} \sin\theta_{a+1}^{(1)} \cos\theta_{a-2}^{(1)} \cos\theta_{a+2}^{(1)} \sin\theta_{a-1}^{(2)} \sin\theta_{a+1}^{(2)} \Big].
\end{aligned}
\end{equation}

This result explicitly demonstrates the topology of information spreading in the quantum circuit. While the state of qubit $a$ after the first layer depends only on its nearest neighbors ($a-1$, $a+1$), the expectation value after the second layer incorporates parameters from next-nearest neighbors ($a-2$, $a+2$). As the circuit depth increases, the local observables become dependent on an exponentially growing neighborhood, which is reminiscent of the Lieb-Robinson bounds for correlation propagation \cite{lieb1972finite} and drives the rapid generation of multi-partite entanglement across the entire chain \cite{amico2008entanglement}.

Finally, substituting these expectation values into the definition of the entanglement distance from Eq.~(\ref{ent dist}), and noting that $\langle Y_a \rangle_2 = 0$, we obtain the complete analytical expression for the bipartite entanglement of the target qubit after two layers
\begin{eqnarray}
    E_2^{ED} = 1 - \langle X_a \rangle_2^2 - \langle Z_a \rangle_2^2 
    = 1 - \Biggl( \sin\theta_a^{(1)} \cos\theta_a^{(2)} \cos\theta_{a+1}^{(2)} \cos\theta_{a-1}^{(2)} \nonumber  \\
    \quad + \sin\theta_a^{(2)} \Bigl[ \cos\theta_a^{(1)} \cos\theta_{a+1}^{(1)} \cos\theta_{a+1}^{(2)} \cos\theta_{a-1}^{(1)} \cos\theta_{a-1}^{(2)} \nonumber  \\
    \qquad - \sin\theta_{a+1}^{(1)} \sin\theta_{a+1}^{(2)} \cos\theta_{a+2}^{(1)} \cos\theta_{a-1}^{(1)} \cos\theta_{a-1}^{(2)} \nonumber  \\
    \qquad - \sin\theta_{a-1}^{(1)} \sin\theta_{a-1}^{(2)} \cos\theta_{a+1}^{(1)} \cos\theta_{a+1}^{(2)} \cos\theta_{a-2}^{(1)} \nonumber  \\
    \qquad + \sin\theta_{a+1}^{(1)} \sin\theta_{a+1}^{(2)} \sin\theta_{a-1}^{(1)} \sin\theta_{a-1}^{(2)} \cos\theta_a^{(1)} \cos\theta_{a+2}^{(1)} \cos\theta_{a-2}^{(1)} \Bigr] \Biggr)^2 \nonumber  \\
    \quad - \left( \cos\theta_a^{(1)} \cos\theta_a^{(2)} - \sin\theta_a^{(1)} \sin\theta_a^{(2)} \cos\theta_{a+1}^{(1)} \cos\theta_{a-1}^{(1)} \right)^2.
\label{ee22}
\end{eqnarray}

So,   the entanglement of a qubit $a$ with the other qubits in
a two-layer multi-qubit variational quantum state depends on the parameters of the $Ry$ gates $\theta_a^{(k)}$, $\theta_{a+1}^{(k)}$, $\theta_{a-1}^{(k)}$, $\theta_{a+2}^{(k)}$, $\theta_{a-2}^{(k)}$, $k=(1,2)$ that  acts on qubits
representing the first- and second-nearest neighbors in the chain topology of the entangled layers.

\section{Quantification of entanglement in variational quantum states with quantum computing}
\label{sec: Simulation}

To verify the analytical expressions obtained for the two-layer case, we performed quantum computing simulations using the \texttt{qiskit\_aer} simulator from the open-source Qiskit framework \cite{Qiskit}. We considered a closed one-dimensional chain of $N=5$ qubits and applied two layers of the parameterized quantum circuit. Instead of using the exact statevector, the expectation values $\langle X_a \rangle$, $\langle Y_a \rangle$, and $\langle Z_a \rangle$ were evaluated from measurement results, using $1024$ shots for each basis. The qubit $a$ was chosen as the central qubit of the chain.

\begin{figure}[htpb]
    \centering
    \scalebox{0.85}{
    \Qcircuit @C=2.2em @R=0.9em @!R { \\
        \nghost{{q}_{0} :  } & \lstick{{q}_{0} :  } & \gate{\mathrm{R_Y}(\theta_0^{(1)})} & \ctrl{1}    & \qw         & \qw         & \qw         & \ctrl{4} \barrier[0em]{4} & \qw \mbox{\makebox[0pt]{\raisebox{4.0em}{$\ket{\psi_1}$}}} & \gate{\mathrm{R_Y}(\theta_0^{(2)})} & \ctrl{1}    & \qw         & \qw         & \qw         & \ctrl{4} \barrier[0em]{4} & \qw \mbox{\makebox[0pt]{\raisebox{4.0em}{$\ket{\psi_2}$}}} & \qw \\
        \nghost{{q}_{1} :  } & \lstick{{q}_{1} :  } & \gate{\mathrm{R_Y}(\theta_1^{(1)})} & \control\qw & \ctrl{1}    & \qw         & \qw         & \qw                       & \qw                                                        & \gate{\mathrm{R_Y}(\theta_1^{(2)})} & \control\qw & \ctrl{1}    & \qw         & \qw         & \qw                       & \qw                                                        & \qw \\
        \nghost{{q}_{2} :  } & \lstick{{q}_{2} :  } & \gate{\mathrm{R_Y}(\theta_2^{(1)})} & \qw         & \control\qw & \ctrl{1}    & \qw         & \qw                       & \qw                                                        & \gate{\mathrm{R_Y}(\theta_2^{(2)})} & \qw         & \control\qw & \ctrl{1}    & \qw         & \qw                       & \qw                                                        & \qw \\
        \nghost{{q}_{3} :  } & \lstick{{q}_{3} :  } & \gate{\mathrm{R_Y}(\theta_3^{(1)})} & \qw         & \qw         & \control\qw & \ctrl{1}    & \qw                       & \qw                                                        & \gate{\mathrm{R_Y}(\theta_3^{(2)})} & \qw         & \qw         & \control\qw & \ctrl{1}    & \qw                       & \qw                                                        & \qw \\
        \nghost{{q}_{4} :  } & \lstick{{q}_{4} :  } & \gate{\mathrm{R_Y}(\theta_4^{(1)})} & \qw         & \qw         & \qw         & \control\qw & \control\qw               & \qw                                                        & \gate{\mathrm{R_Y}(\theta_4^{(2)})} & \qw         & \qw         & \qw         & \control\qw & \control\qw               & \qw                                                        & \qw \\
    \\ }}
    \vspace{0.2cm}
    \caption{Quantum circuit diagram for a 5-qubit closed 1D chain ($N=5$) with two layers ($n=2$). Each layer applies local $R_Y$ rotations followed by entangling $CZ$ gates between adjacent pairs, applied sequentially. The control line spanning from $q_0$ to $q_4$ explicitly denotes the periodic boundary connection that closes the chain. The state of the system after each layer is denoted by $\ket{\psi_n}$. In our simulations, $q_2$ is designated as the target qubit $a$.}
    \label{fig:5qubit_circuit}
\end{figure}
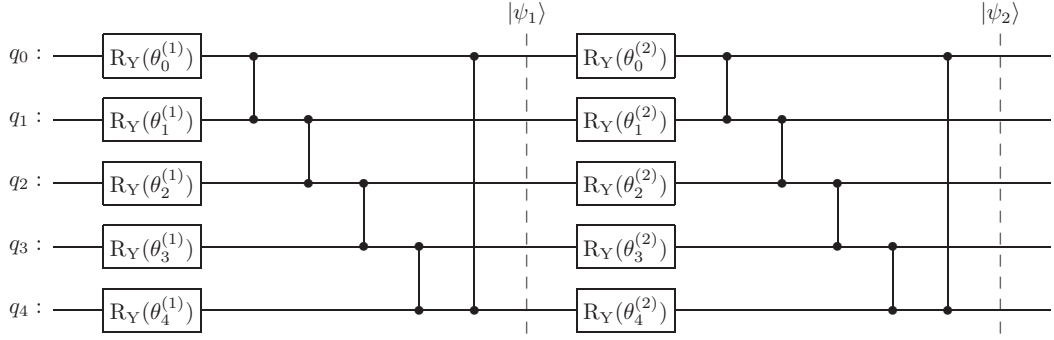

\begin{figure}[htpb]

    \centering

    \includegraphics[width=0.6\textwidth]{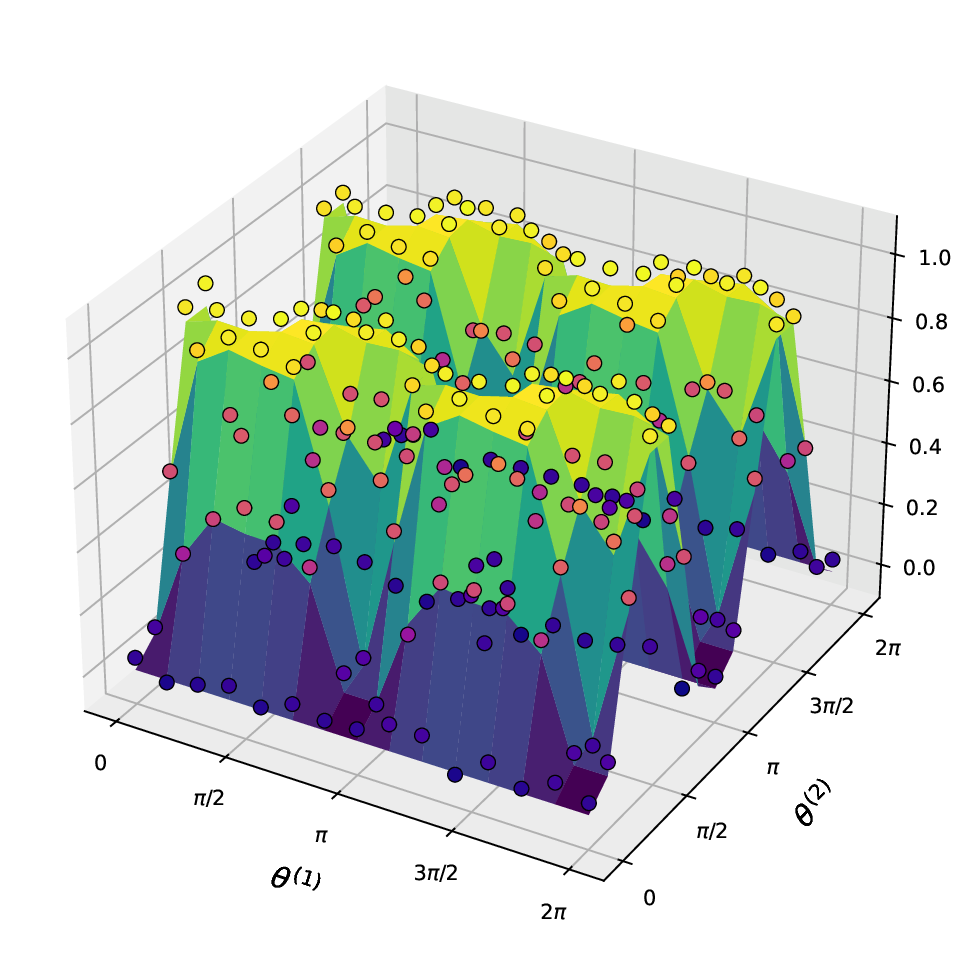}

    \caption{Surface plot of the entanglement distance $E_a^{ED}$ as a function of independent layer angles $\theta^{(1)}$ and $\theta^{(2)}$. The semi-transparent surface represents the analytical derivation, while the distinct points are obtained via simulated quantum measurements. The landscape demonstrates a rich periodic structure dependent on the interplay between consecutive layers.\\
    $ E_2^{ED} = 1 - \left( \sin\theta^{(1)} \cos^3\theta^{(2)} + \sin\theta^{(2)} \left[ \cos^3\theta^{(1)} \cos^2\theta^{(2)} - 2 \sin\theta^{(1)} \cos^2\theta^{(1)} \sin\theta^{(2)} \cos\theta^{(2)} \right. \right. $\\
$\left. \left. + \sin^2\theta^{(1)} \cos^3\theta^{(1)} \sin^2\theta^{(2)} \right] \right)^2
- \left( \cos\theta^{(1)} \cos\theta^{(2)} - \sin\theta^{(1)} \cos^2\theta^{(1)} \sin\theta^{(2)} \right)^2 $ }

    \label{fig:3d_layers}

\end{figure}

To further explore the expressivity of the parameterized quantum circuit ansatz \cite{sim2019expressibility}, we relax the condition of identical angles. In Figure~\ref{fig:3d_layers}, we plot the entanglement distance surface where the rotation angle is uniform within each layer, but differs between layers ($\theta^{(1)}$ for Layer 1 and $\theta^{(2)}$ for Layer 2).

\begin{figure}[htpb]

    \centering

    \includegraphics[width=0.6\textwidth]{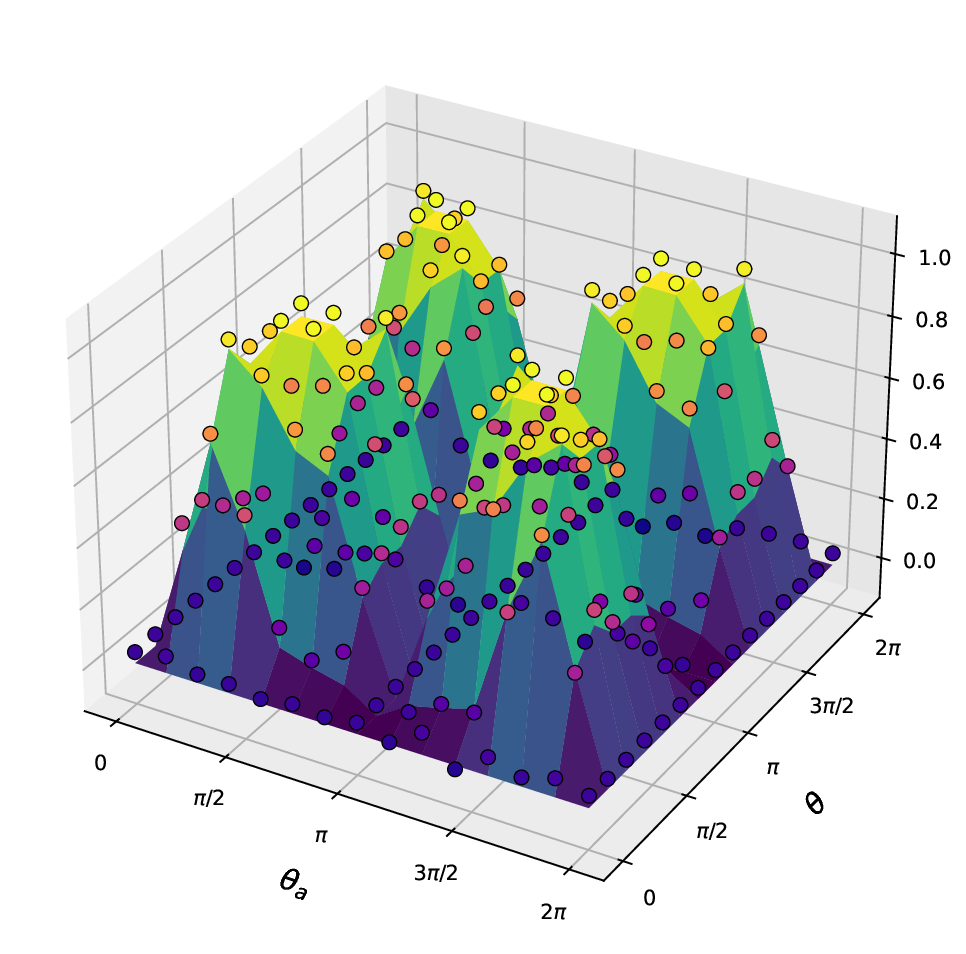}

    \caption{Entanglement distance surface showing the interplay between the target qubit's local rotation angle $\theta_a$ and the surrounding environment's rotation angle $\theta$. The perfect alignment of the simulated data points (quantum computing) with the underlying analytical surface confirms the validity of the derived topological correlation spreading equations. \\
    $E_2^{ED} = 1 - \left( \sin\theta_a \cos\theta_a \cos^2\theta + \sin\theta_a \left[ \cos\theta_a \cos^4\theta - 2 \sin^2\theta \cos^3\theta
+ \sin^4\theta \cos\theta_a \cos^2\theta \right] \right)^2 - \left( \cos^2\theta_a - \sin^2\theta_a \cos^2\theta \right)^2$}

    \label{fig:3d_target_env}

\end{figure}

Finally, we analyze a scenario where the target qubit $a$ is subjected to local rotations by an angle $\theta_a$, while all other qubits in the ``environment'' are rotated by a uniform angle $\theta$. This topology models the sensitivity of the target qubit's entanglement to local control versus global environmental operations. The resulting correlation landscape is shown in Figure~\ref{fig:3d_target_env}.

In all cases, the numerically simulated data points, despite containing statistical shot noise typical for quantum computing protocols, precisely follow the analytically derived continuous surfaces. This confirms that the recursive methodology correctly captures the exact topological dependencies and multi-partite entanglement dynamics generated by the $CZ$ gates over multiple layers.

\section{Conclusions}
\label{sec: Conclusions}

In this work, we study the entanglement distance of quantum states generated by parameterized quantum circuits with $R_Y$ rotations and $CZ$ gates.  For the two-qubit case, we obtain analytical expressions that are valid for an arbitrary number of circuit layers, allowing us to track how local observables change from layer to layer (\ref{eq:E_layer_n}).

We also extend this analysis to a multi-qubit system arranged in a closed one-dimensional chain. For the two-layer circuit, we obtain explicit analytical expressions for the expectation values. These results show that the expectation values of a given qubit depend on the parameters of gates acting on its neighboring qubits and that this dependence grows as the number of layers increases, reflecting the spreading of correlations along the chain~\cite{DeSimone2025Directed}. On the basis of the obtained results we find the complete analytical expression for the bipartite entanglement of the target qubit after two layers (\ref{ee22}).  We obtain that the entanglement of a qubit with the other qubits in a two-layer variational quantum state depends on the parameters of the gates acting on qubits representing the first- and second-nearest neighbors in the chain topology of entangled layers.

 We also performed numerical simulations using quantum programming tools. We considered different parameter settings by varying the rotation angles in different layers. In all cases, the numerical results are in good agreement with the analytical predictions (see Figs. 2-4, 6, 7). The small differences are due to statistical errors caused by the finite number of measurement shots.

These results provide a useful analytical description of how correlations and entanglement develop in parameterized quantum circuits. The proposed approach can be applied in future studies, for example, to investigate the effect of noise on entanglement distance or to consider other multi-qubit circuit structures relevant for near-term quantum devices.

\bibliographystyle{unsrt}

\bibliography{bibliography}

\end{document}